# Ray-Space Motion Compensation for Lenslet Plenoptic Video Coding

Thuc Nguyen Huu, Vinh Van Duong, Jonghoon Yim, and Byeungwoo Jeon

*Abstract*—Plenoptic images and videos bearing rich information demand a tremendous amount of data storage and high transmission cost. While there has been much study on plenoptic image coding, investigations into plenoptic video coding have been very limited. We investigate the motion compensation for plenoptic video coding from a slightly different perspective by looking at the problem in the ray-space domain instead of in the conventional pixel domain. Here, we develop a novel motion compensation scheme for lenslet video under two sub-cases of ray-space motion, that is, integer ray-space motion and fractional ray-space motion. The proposed new scheme of light field motion-compensated prediction is designed such that it can be easily integrated into well-known video coding techniques such as HEVC. Experimental results compared to relevant existing methods have shown remarkable compression efficiency with an average gain of 19.63% and a peak gain of 29.1%.

*Index Terms*—Plenoptic video, coding, compression, light field video, ray-space motion model, motion compensation.

## I. Introduction

Since G. Lipman's inspiring idea [1], integral photography makes much more practical sense these days due to breakthroughs both in camera (e.g., unfocused [6] and focused [7] plenoptic cameras) and display (e.g., recent industry products like [8] [9]) technologies. The revolutionary concept of recording and reconstructing rays in 3D space has greatly influenced the evolution of 3D imaging, as well as computational photography (e.g., digital re-focusing [2], dense depth/disparity extraction [3]). The ray information in the space is modeled by a 5D plenoptic function [4] (or 4D light field [5]), which describes the ray's radiance at every position along every direction in the scene. Plenoptic cameras using the lenslet array (a.k.a. microlens array) map plenoptic data into a 2D image sensor, forming a plenoptic image represented in the lenslet format [6]. The plenoptic image can also be organized in the multiview format where each view (picture) is a collection of rays of the same viewpoint. While the usefulness of such high-dimensional data is undeniable, it faces challenges related to its huge data storage and high transmission cost, even when using current advanced technologies. For example, an RGB lenslet image captured using the Lytro Illum camera with a spatial resolution of 7728 x 5368 consumes approximately 118 megabytes in cases without compression (assume 8 bits per pixel). The challenge becomes even more serious with plenoptic video, where an additional dimension of time is added. Therefore, compression of plenoptic images and video has brought up an important, practical issue, promoting considerable research interest from both industry and academia.

For plenoptic image coding, there are many proposed solutions, which can be classified as either (1) a transform-based coding approach or (2) a prediction-based coding approach. The transform-based coding approach decorrelates the plenoptic data using well-known transforms, such as 3D DCT [10], KLT [11], DWT [12], etc. For example, 4D DCT was used for the multiview format [14]. The prediction-based coding approach aims to efficiently predict the signal by using well-known prediction tools. For example, for the lenslet format, several approaches can be used, including the intra bock copy (IBC) tool [14], intra displacement vector prediction [15], self-similarity (SS) prediction [16], locally linear embedding (LLE) prediction [17], and reshaping and boundary matching (BMP) prediction [18] [19], which basically finds or computes the prediction that has the best spatial matching within the current lenslet image. With respect to the multiview format, a few examples noted here are pseudo video sequence (PVS)-based methods [20] [21], which scan 2D view arrays following particular scan-orders (for example, zigzag, spiral, or raster order); inter-view prediction methods [22] [23], which make use of the full 2D view parallax; and view synthesis prediction in [24] [25] [26]. These transform-based and prediction-based coding approaches are independent of each other and can be combined to form a hybrid approach. In fact, most of the predictive coding methods reviewed above improve the intra/inter prediction tools in coding methods, such as HEVC, while employing the conventional block-based DCT in the rest of the compression system.

For plenoptic video coding, early studies investigated the use of existing plenoptic image coding tools that are also applied to the video case. For instance, with respect to the lenslet format, the BMP tool [27] was reused for videos captured by a focused plenoptic camera. With respect to the multiview format, [28]

Thuc Nguyen Huu, Vinh Van Duong, Jonghoon Yim, and Byeungwoo Jeon (corresponding author) are with Digital Media Laboratory (DMLAB), Department of Electrical and Computer Engineering, Sungkyunkwan University, 2066 Seoburo, Jangan-gu, Suwon, 16410, KOREA (email: thuckechsu, duongvinh, yjh6522, bjeon@skku.edu).

This research was supported by the Basic Science Research Program through the National Research Foundation of Korea (NRF) funded by the Ministry of Science and ICT (NRF-2020R1A2C2007673).

Table I. Notations

| Symbol | Description |
| --- | --- |
| $\mathbf{r} = \begin{bmatrix} u & v & s & t \end{bmatrix}^T$ | The light ray coordinate of the 4D light field. |
| $l(\mathbf{r})$ or $l(u,v,s,t)$ | The 4D light field function using vector notation or scalar notation. |
| $p_s, p_t$ | The horizontal and vertical distances of microlenses in the microlens array (MLA). |
| $P_x, P_y$ | The horizontal and vertical distances of micro-images in the lenslet image. |
| $i(\mathbf{x})$ | The image sensor function representing the pixel brightness at position $\mathbf{x} = \begin{bmatrix} x & y \end{bmatrix}^T$. |
| $\omega_\alpha(m)$ | 1D filter coefficients used for the interpolation at fractional position $\alpha$. |

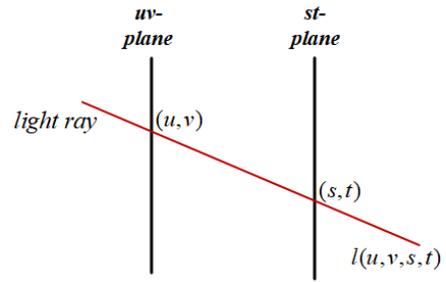

Fig. 1. Light field representation using two-parallel-plane ray parameterization. [6]

[29] [30] [31] [32] proposed inter-view prediction structures while re-using the conventional temporal prediction structure (the group of pictures [33]) and applying them to MV-HEVC [34]. Apart from those, the macropixel-constrained collocated position (MCP) search [35] technique proposed a motion searching pattern for lenslet-based plenoptic video. The MCP search outperforms the test zone (TZ) search [36], which is widely used in the HEVC encoder, while achieving comparable performance to the full search. Besides, [37] proposed an interesting coding framework called the steered mixture-of-experts (SMoE), which approximates the plenoptic data by using a set of kernels. Here, a kernel can be understood as a centroid point (similar to one used in vector quantization [38]), which groups the light rays arriving at a certain region from surrounding angles. The SMoE framework is reported to achieve better coding performance than the MV-HEVC approach.

Notably, international standardization bodies have also paid attention to plenoptic/light field compression. That is, the Joint Photographic Experts Group (JPEG) has started the JPEG Pleno [39] project, which addresses emerging issues of plenoptic image representation and coding. It has organized two grand challenge events concerning plenoptic images: one at the IEEE International Conference on Image on Multimedia and Expo (ICME) in 2016 [40] and the other at the IEEE International Conference on Image Processing (ICIP) in 2017 [41]. Besides, the Moving Picture Experts Group (MPEG) is currently working on plenoptic video for immersive applications, referred to as 6-degrees-of-freedom (6-DoF) VR/AR, 3D imaging and display, and 3D graphics-based point cloud. These working items are grouped into *MPEG-I* where "*I*" stands for "*immersive*" [42] [43]. In *MPEG-I*, plenoptic video coding is being actively discussed, in which suitable plenoptic representation/formatting as well as common test conditions are under construction [44].

It can be safely said that plenoptic video coding is still in its early stages. It has yet to fully exploit redundancies in the temporal domain, meaning it has yet to achieve the maximum possible compression efficiency. To reduce the temporal redundancy, traditional motion compensation that has been widely used in current video coding standards (e.g., H.264/AVC, H.265/HEVC, H.266/VVC) can also be used to find motion vectors, associating similar blocks among plenoptic pictures. However, it should be further noted that unlike pictures of conventional 2D video, plenoptic video data is a collection of ray radiances in space over time. Therefore, its motion is tightly associated with the rays in the 3D space (we denote this as *ray-space motion*). In Section III, the ray-space motion is studied in detail in the context of a plenoptic camera, by which a new motion compensation scheme of lenslet images can be designed. Following this, we propose a new motion-compensated prediction directly addressing the ray-space motion and apply it to lenslet-based plenoptic video coding. Compared to the traditional motion compensation of HEVC, the proposed method achieves a remarkable compression gain of up to 29.1% when implemented in HEVC and applied to plenoptic video.

The paper is organized as follows. Section II explains fundamental background information. We address the concept of the proposed ray-space motion in the context of plenoptic cameras in Section III. In Section IV, based on the ray-space motion concept, we derive the motion compensation model of lenslet images for two specific cases: integer and fractional ray-space motion. We also describe how our motion compensation model can be implemented on top of existing coding schemes, such as HEVC, to realize the ray-space motion-compensated prediction. Finally, we discuss our experimental results in Section VI and then conclude the paper in Section VII.

## II. Background

### A. Light field representation

We follow the two-parallel-plane ray parameterization [6], which records the ray intersection with two parallel planes — a *uv*-plane and an *st*-plane — as shown in Fig. 1. Using the parameterization, the radiance (or intensity) of light rays is represented as a 4D signal $l(u,v,s,t)$, called the *light field* (LF), which can describe the ray radiance at every position along every direction in the space. The 4D light field is a practical realization of the general 5D plenoptic function [4] under the assumption that the radiance of the light ray does not change along its direction. One easy interpretation of the light field is to consider a set of cameras on the *uv*-plane with their focal plane on the *st*-plane, where each view stores information of light rays departing from the same viewpoint *uv*. Here, each view is considered as a 2D slice of the light field $l_{uv}(s,t)$ for a specific *uv*. This is referred to as the multiview format, which is shown in Fig. 2(a). Alternatively, the LF data can also be represented as an array of $l_{st}(u,v)$ images for a specific *st*, as shown in Fig. 2(b). This is referred to as the lenslet format. In this configuration, a micro-image stores information of all light rays passing through its corresponding microlens at the *st* position onto the image sensor.

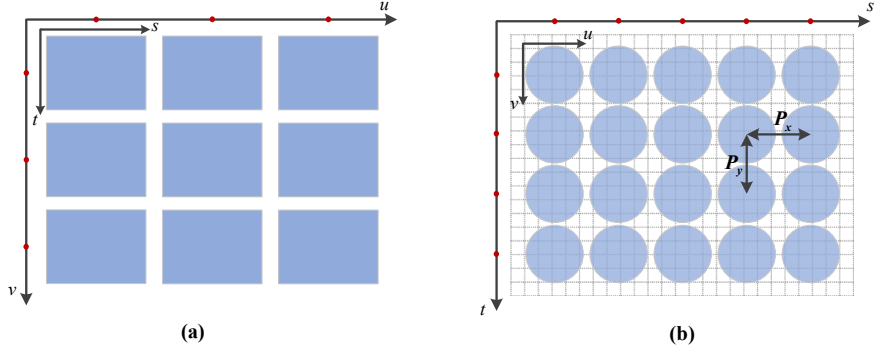

Fig. 2. (a) The multiview format with an array of $l_{uv}(s,t)$ images (or so-called views). (b) The lenslet format with an array of micro-images $l_{st}(u,v)$; each micro-image (drawn in a circular shape) stores information of all light rays passing through a corresponding microlens at the *st* position.

The light field can be represented either in the multiview format or the lenslet format. Generally, conversion between the two formats is invertible by a simple data re-arrangement and a slice operator [5]. When a lenslet image is obtained by conversion from the multiview format, some researchers have [47] [48] [49] referred to it as a '*synthetic/rectified lenslet*' image to differentiate it from the *raw lenslet* output of the plenoptic camera. It is also noted here that there exists another lenslet format obtained from the *focused plenoptic camera* (e.g., Raytrix) [7], known as the *focused-lenslet format*. In this case, each micro-image is not a 2D slice $l_{st}(u,v)$ of the light field; thus, conversion between the multiview and the focused-lenslet format is non-invertible. Some early investigations [44] have shown that multiview coding is much more efficient than focused-lenslet coding. Unless the main aim is for camera storage, focused-lenslet coding is seen to be less widely applicable than multiview coding. In this paper, therefore, the plenoptic camera (unless otherwise specified) is understood as following the unfocused plenoptic camera model.

### B. Lenslet plenoptic video coding

As reported in [48] [49], choosing a proper LF representation greatly impacts the overall compression performance of light field coding. Besides, it should also consider practical convenience in terms of capturing, processing, transmitting, and applying [50] LF data. In this paper, we adopt the lenslet format for light field video compression due to its wide usages [51], ranging from camera storage (e.g., raw lenslet images are captured by plenoptic cameras [6] [7]) to integral display [52]. Additionally, the works by MPEG-I [44] [53] have also shown evidence that lenslet coding is much more efficient than multiview coding. Considering the easy nature of format conversion by using a simple re-arrangement and slice operator, one could use the lenslet format in coding to achieve greater compression performance.

## III. RAY-SPACE MOTION MODEL IN PLENOPTIC CAMERA

### A. Basic configuration

Fig. 3 shows the basic configuration of the *standard plenoptic camera* [6] (also known as the *unfocused plenoptic camera*), which captures LF data in the lenslet format. It has a microlens array (MLA) between a main lens and an image sensor. $f_{lens}$ and $f_\mu$ denote the distances of MLA from the main lens and the image sensor, respectively. The 4D light field $l(u,v,s,t)$ is formed by collecting light rays connecting the main lens (at the *uv*-plane) and MLA (at the *st*-plane). The origin of each plane is the intersection of the optical axis with the corresponding plane where the optical axis is a straight line that is perpendicular to the main lens at its center. Moreover, the main lens is assumed to follow the thin lens model while the MLA is treated as the pinhole array model where the centers of microlenses are placed following a rectangular pattern of spacing $(p_s, p_t)$ along horizontal and vertical direction, which is referred to a rectangular MLA structure. The microlenses can also be arranged according to a hexagonal pattern to maximize the fill factor of the sensor. This is called the hexagonal MLA structure, and it generates micro-images in a hexagonal structure. In this paper, we develop our scheme by assuming rectangular MLA only, since re-mapping from the hexagonal structure to the rectangular one is quite straightforward [66] [77]. It is also noted that re-mapping is widely used to extract a desired view of the light field [55] [66]. Re-mapping gives us the flexibility to treat the lenslet data uniformly, irrespective of whether it was captured by rectangular or hexagonal MLA. This flexibility is extended even to the synthetic lenslet (converted from the multiview format).

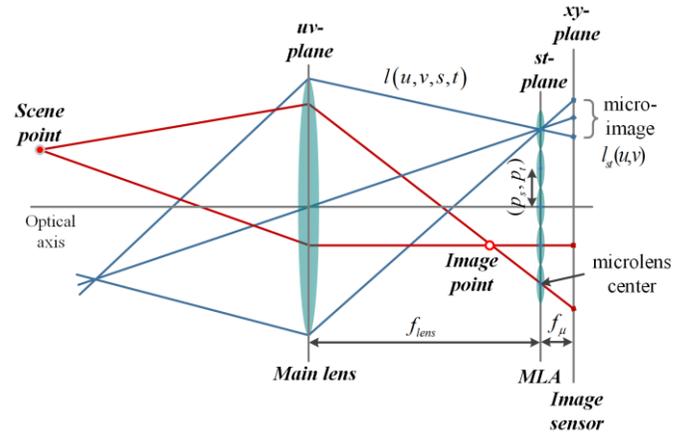

Fig. 3. Plenoptic camera configuration.

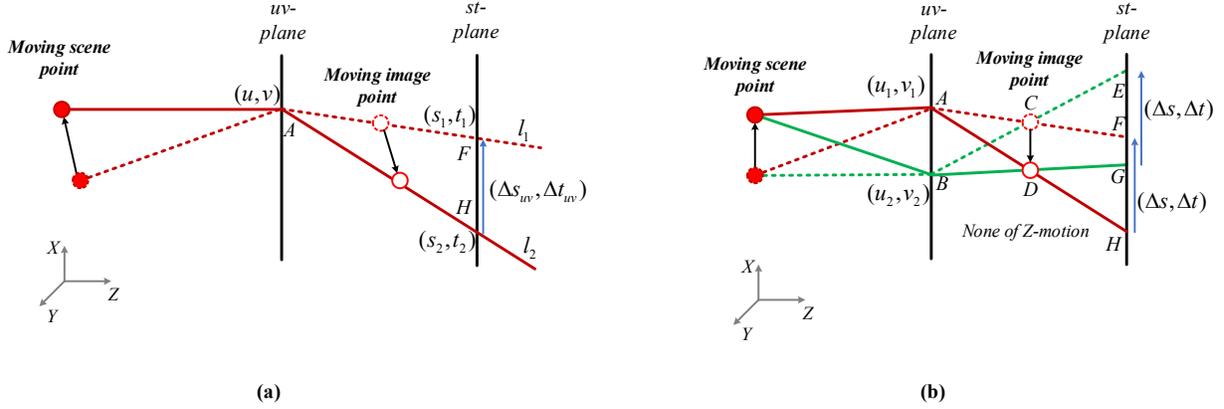

Fig. 4. The concept of the proposed ray-space motion model. Light rays at different time instants are shown by solid and dotted lines. (a) The general ray-space motion model and (b) a simplified ray-space motion model assuming no motion in the Z-axis.

*B. Scene point motion analysis*

Let us assume that a scene points in Fig. 3 satisfies the Lambertian [54] surface property. A point on the Lambertian surface appears equally bright from all viewing directions. In Fig. 3, the optical path from a scene point in front of the plenoptic camera can be traced as follows. First, the light rays emitted from a scene point converge at a point through the main lens; we call this point the *image point* (note that the term '*image point*' here does not refer to a pixel recorded by the image sensor, but to the converging point of light rays by the main lens). By further tracing the light rays to reach the image sensor, we find a complete optical path of rays emanating from a particular scene point to the image sensor through the center of a microlens. Due to the Lambertian surface property, a scene point, irrespective of direction, emits equally bright light rays (drawn as red lines in Fig. 3). These rays are eventually recorded by the image sensor at multiple locations throughout the sensor plane. This could explain the high similarity that exists among micro-images. In fact, many state-of-the-art coding solutions of plenoptic images in the lenslet format [16] [17] [19] have made use of this similarity. However, they have noted the similarity only in the spatial domain (but not in the temporal domain), simply because the solutions were originally developed just for image coding. In this paper, we would like to take fuller advantage of the similarity that exists in the temporal sequence of lenslet images (that is, a video) by investigating a motion model for the plenoptic video.

If a scene point moves, its image point also moves according to a linear mapping relation by the main lens [55]. Note that the light rays coming to an image point are directed to multiple locations of the image sensor due to microlenses, thereby forming a group of pixels. The movement of an image point that is created by the object movement will then also shift the group in the sensor plane, which is eventually recorded by the image sensor. This may sound similar to the *scene flow* (three-dimensional motion field of scene points [56]) studied by some researchers in association with the multiview camera [57] or plenoptic camera [58] [59]. However, the effect of the scene point motion on the image sensor especially related to the plenoptic camera has not been investigated in detail yet, and its study goal is different from ours. The scene flow problem associated with the plenoptic camera aims to find the 3D motion field given the plenoptic video data. Alternatively, in our problem, we aim to study the changes of lenslet images due to scene point motion, which later allows us to establish a new motion compensation model for a temporal sequence of lenslet images. In this regard, a new motion-compensated prediction method can be designed to replace the traditional motion compensation used in existing video coding techniques for lenslet plenoptic video coding.

*C. Model for scene point motion to ray-space motion*

To design the motion compensation model, we introduce the concept of *ray-space motion* (the ray motion in 3D space) and investigate the association of the scene point motion with the corresponding changes in lenslet images. Let us consider an image point in the space between the main lens and MLA. If we draw two light rays $l_1$ and $l_2$ connecting a viewpoint $(u,v)$ with the two image points corresponding to before ($l_1$ drawn in dotted line) and after ($l_2$ drawn in solid line) scene point motion, respectively, we can see the displacement vector of the light rays $(\Delta s_{uv}, \Delta t_{uv})$ at the *st*-plane, as shown in Fig. 4(a). Here, we define the ray displacement vector (or ray motion vector) as a two-dimensional vector indicating an offset from the *st* coordinates of $l_2$, the current light field image to the *st* coordinates of $l_1$, the reference light field image as defined below:

$$\begin{cases} \Delta s_{uv} = s_1 - s_2 \\ \Delta t_{uv} = t_1 - t_2 \end{cases}. \quad (1)$$

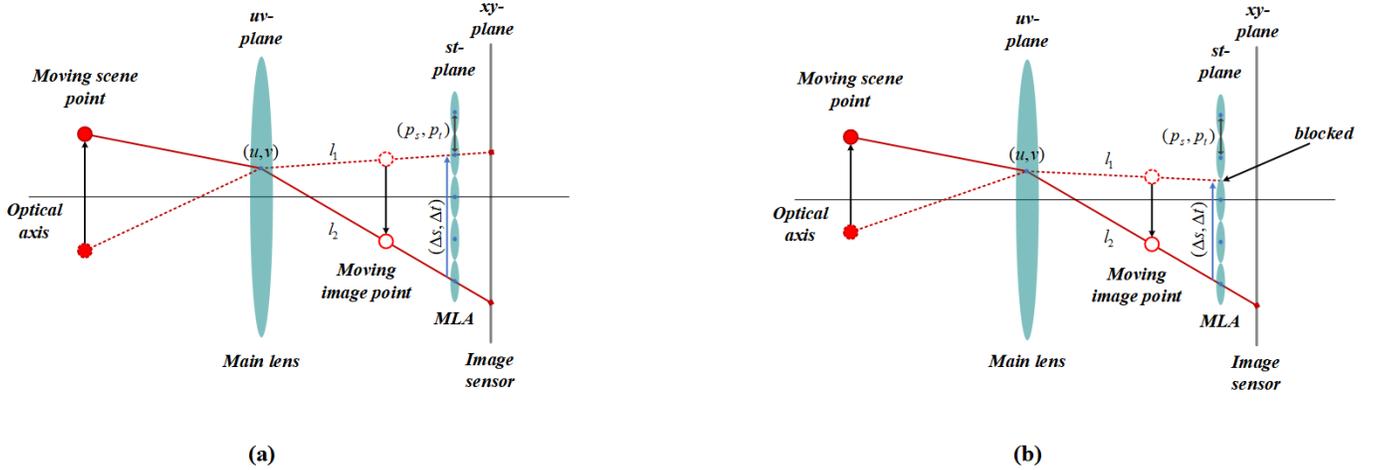

Fig. 5. Two specific cases of the ray-space motion model: (a) integer ray-space motion and (b) fractional ray-space motion.

To avoid any confusion, we note here that our ray motion vector is different from the conventional motion vector. The former indicates the relationship of *st* locations on the MLA plane, whereas the latter indicates the relationship of pixel locations *xy* on the sensor plane. The ray displacement vector generally depends on a specific *uv* position. Because the light rays $l_1$ and $l_2$ are emitted from the same scene point (but at different time instants), the radiances of light rays $l_1$ and $l_2$ are assumed to be equal, thus implying that:

$$l_2(u,v,s,t) = l_1(u,v,s+\Delta s_{uv}, t+\Delta t_{uv}). \quad (2)$$

This relationship helps to predict the light field $l_2$ from its reference light field $l_1$ by using the ray displacement vector $(\Delta s_{uv}, \Delta t_{uv})$ representing the scene motion. We denote Eq. (2) as the general ray-space motion model. It can be further simplified by assuming no (or negligible) motion in the Z-axis (called *Z-motion*) of the scene point, as shown in Fig. 4(b); here, the optical axis is treated as the Z-axis. This assumption is mostly valid unless the frame rate of the light field video is too small compared to the speed of motion in the Z-direction. The assumption of no Z-motion of a scene point would also mean no Z-motion of an image point, which can be verified using the Gaussian thin lens equation [60]. Then, in the context of the proposed ray-space motion, the ray displacement vector $(\Delta s_{uv}, \Delta t_{uv})$ resulting from the motion of the same scene point is constant for all *uv*. This means that every viewpoint observes the same displacement vector. According to Fig. 4(b), the length *GE* should be equal to *HF* (refer to Appendix A for more details). Therefore, we obtain a simplified ray-space motion model:

$$l_2(u,v,s,t) = l_1(u,v,s+\Delta s, t+\Delta t), \quad (3)$$

where $(\Delta s, \Delta t)$ is a constant ray displacement vector. It becomes handier later to represent this using its vector notation:

$$l_2(\mathbf{r}) = l_1\left(\mathbf{r} + \begin{bmatrix} 0 & 0 & \Delta s & \Delta t \end{bmatrix}^T\right), \quad (4)$$

where $\mathbf{r} = \begin{bmatrix} u & v & s & t \end{bmatrix}^T$ is a light ray coordinate of the light field. An advantage of using the simplified ray-space motion model is clear since the vector $(\Delta s, \Delta t)$ is all that is needed to model the motion of a scene point in the ray-space. Later, we will show that the ray displacement vector can be used to efficiently describe the motion compensation model of a lenslet plenoptic video. From now on, we only consider the simplified ray-space motion model given that the scene point has no z-motion (or has only negligible z-motion).

*D. Integer and fractional ray-space motion*

Related to the image sensor, we consider two specific cases of the ray-space motion: *the integer* and *the fractional ray-space motions*. Under the pinhole assumption of the microlens, an incoming light ray is thought to go through the MLA only at the center of a microlens. The center position of each microlens is referred to by the integer MLA position, while the off-center position is referred to by the fractional MLA position. These notations are borrowed from the traditional integer and fractional pixel positions [61]. Under the ray-space motion model with the ray displacement vector $(\Delta s, \Delta t)$, the integer ray-space motion describes the case when the light ray $l_1$ reaches the image sensor through the centers of microlenses, that is, the integer MLA positions, as shown in Fig. 5(a). This case is represented by the integer ray-space motion. By the same token, the fractional ray-space motion refers to the case when the light ray $l_1$ at the first incident is blocked by MLA due to its arrival at a fractional MLA position; after the motion, the ray $l_2$ reaches at the center of a microlens at a later time instant, as shown in Fig. 5(b). Due to the blocking of the fractional ray at the first time instant, difficulty arises in finding matching rays in the reference light field $l_1$ corresponding to the moving scene point in the motion estimation process.

Under the assumption of rectangular MLA (as mentioned in sub-section III-A), the centers of microlenses follow a rectangular pattern with horizontal and vertical distances of $(p_s, p_t)$ at the *st*-plane, as shown in Fig. 3. The coordinate of light rays going through integer MLA positions can be denoted as $\mathbf{r}_{int}$ as below where $\mathbf{Z}$ is an integer set:

$$\mathbf{r}_{int} = \begin{bmatrix} u & v & k_s p_s & k_t p_t \end{bmatrix}^T; \quad k_s, k_t \in \mathbf{Z}. \tag{5}$$

In the case of the integer ray-space motion shown in Fig. 5(a), the ray displacement vector $(\Delta s, \Delta t)$ should satisfy the integer ray-space motion, described as follows:

$$l_2(\mathbf{r}_{int}) = l_1\left(\mathbf{r}_{int} + \begin{bmatrix} 0 & 0 & \Delta k_s p_s & \Delta k_t p_t \end{bmatrix}^T\right); \Delta k_s, \Delta k_t \in \mathbf{Z}, \tag{6}$$

where $(\Delta k_s, \Delta k_t)$ is the integer ray displacement vector. In the case of the fractional ray-space motion shown in Fig. 5(b), the fractional ray displacement vector $(\Delta s, \Delta t)$ can represent the fractional ray-space motion as:

$$l_2(\mathbf{r}_{int}) = l_1\left(\mathbf{r}_{int} + \begin{bmatrix} 0 \\ 0 \\ (\Delta k_s + \alpha) p_s \\ (\Delta k_t + \beta) p_t \end{bmatrix}\right); \quad 0 < \alpha, \beta < 1. \tag{7}$$

The ray displacement is decomposed into its integer part $(\Delta k_s, \Delta k_t)$ and its fractional part $(\alpha, \beta)$ Without loss of generality, we assume $0 < \alpha, \beta < 1$.

IV. MOTION COMPENSATION FOR RAY-SPACE MOTION MODEL

This section derives the motion compensation model of the lenslet video based on the integer and fractional ray-space motions defined in the previous section.

*A. Derivation of the integer ray-space motion*

Eq. (6) specifies a constraint of the integer ray-space motion in which both light rays, before and after motion, reach the image sensor. To derive a motion compensation model, we first introduce a light field projection [66] that maps a light ray coordinate $\mathbf{r} = \begin{bmatrix} u & v & s & t \end{bmatrix}^T$ into an image sensor coordinate $\mathbf{x} = \begin{bmatrix} x & y \end{bmatrix}^T$:

$$\mathbf{T}: \mathbf{r} \to \mathbf{x}$$
$$\mathbf{T} = \begin{bmatrix} -F & 0 & 1+F & 0 \\ 0 & -F & 0 & 1+F \end{bmatrix}, \tag{8}$$

where $F = \dfrac{f_\mu}{f_{lens}}$ satisfies the F-number matching condition [6] from the plenoptic camera design. The pixel value at an arbitrary image sensor location $\mathbf{x}$, denoted by $i(\mathbf{x})$, can be related to the light field value by noting the linear transformation:

$$\mathbf{x} = \mathbf{T}\mathbf{r}_{int}. \tag{9}$$

That is, the light field incoming to an integer MLA position $\mathbf{r}_{int}$ is recorded by the image sensor at $\mathbf{x}$ as:

$$l(\mathbf{r}_{int}) = i(\mathbf{T}\mathbf{r}_{int}). \tag{10}$$

Eq. (6), which constrains the integer ray-space motion, can be equivalently written in terms of image sensor values as:

$$i_2(\mathbf{T}\mathbf{r}_{int}) = i_1\left(\mathbf{T}\left(\mathbf{r}_{int} + \begin{bmatrix} 0 & 0 & \Delta k_s p_s & \Delta k_t p_t \end{bmatrix}^T\right)\right)$$
$$= i_1\left(\mathbf{T}\mathbf{r}_{int} + \mathbf{T}\begin{bmatrix} 0 & 0 & \Delta k_s p_s & \Delta k_t p_t \end{bmatrix}^T\right) \tag{11}$$
$$= i_1\left(\mathbf{T}\mathbf{r}_{int} + (1+F)\begin{bmatrix} \Delta k_s p_s \\ \Delta k_t p_t \end{bmatrix}\right),$$

where $i_1(.)$ and $i_2(.)$ are corresponding values of the image sensor for the light field $l_1(.)$ and $l_2(.)$, respectively. Since we assume $l_2(.)$ and $l_1(.)$ to be the current and reference light fields, respectively, the image sensor functions $i_2(.)$ and $i_1(.)$ can be naturally interpreted as the current and reference lenslet images, respectively. Finally, the motion compensation model under the integer ray-space motion is obtained by letting $\mathbf{x} = \mathbf{T}\mathbf{r}_{int}$:

$$i_2(\mathbf{x}) = i_1\left(\mathbf{x} + \begin{bmatrix} \Delta k_s P_x \\ \Delta k_t P_y \end{bmatrix}\right). \tag{12}$$

Here, we denote the micro-image distances $P_x, P_y$ at the *xy*-plane in the horizontal and vertical directions as shown in Fig. 2(b). $P_x$ and $P_y$ are computed as:

$$\begin{cases} P_x = (1+F) p_s \\ P_y = (1+F) p_t \end{cases}. \tag{13}$$

In practice, these micro-image distances are already known,

even for a raw lenslet or synthetic images (converted from the multiview format). In the case of raw lenslet images, they are usually provided by the camera manufacturer or obtained by the white image calibration [55] [66] [67]. In the case of synthetic lenslet images, they are the angular resolutions of the 4D light field if the micro-images are closely arranged together.

### B. Derivation of the fractional ray-space motion

Eq. (7) gives the constraint for the fractional ray-space motion. However, unlike the previous case, the light ray on the right-hand side of Eq. (7) cannot reach the image sensor. Therefore, we cannot directly apply the light field projection technique to this case. To solve the problem, we first re-write Eq. (7) as:

$$l_2(\mathbf{r}_{int}) = l_1\left(\mathbf{r}_{int}^* + \begin{bmatrix} 0 & 0 & \alpha p_s & \beta p_t \end{bmatrix}^T\right) \quad (14)$$
$$= l_1(\mathbf{r}^*),$$

where $\mathbf{r}_{int}^* = \begin{bmatrix} u & v & (k_s + \Delta k_s)p_s & (k_t + \Delta k_t)p_t \end{bmatrix}^T$ and $\mathbf{r}^* = \mathbf{r}_{int}^* + \begin{bmatrix} 0 & 0 & \alpha p_s & \beta p_t \end{bmatrix}^T$. It is noted that the light field $l_1(\mathbf{r}^*)$ is not recorded by the image sensor and is thus undefined. In motion estimation/compensation, one can still interpolate it out using available neighboring light field data $l_1(\mathbf{r}_{int}^*)$. The interpolation is closely related to *the light field rendering* [5], where one can render or reconstruct light rays that are not located on the sampling grid. Notice that the light ray coordinate $\mathbf{r}^*$ is not located exactly on the sampling grid at the *st*-plane. Instead, it is located on the fractional MLA position specified by parameters $(\alpha, \beta)$. Therefore, a 2D interpolation in the *st*-space is required. The 2D interpolation can be easily accomplished by an FIR filter of size $(2M+1) \times (2N+1)$:

$$l_1(\mathbf{r}^*) = \sum_{m=-M}^{M} \sum_{n=-N}^{N} \omega_{\alpha,\beta}(m,n) l_1\left(\mathbf{r}_{int}^* + \begin{bmatrix} 0 \\ 0 \\ mp_s \\ np_t \end{bmatrix}\right), \quad (15)$$

where $\omega_{\alpha,\beta}$ is an interpolation filter coefficient for the fractional position $(\alpha, \beta)$. The closed form of $\omega_{\alpha,\beta}$ will be discussed in the next section. Combining Eq. (15) with Eq. (14), $l_2(\mathbf{r}_{int})$ is shown to be equal to:

$$l_2(\mathbf{r}_{int}) = \sum_{m=-M}^{M} \sum_{n=-N}^{N} \omega_{\alpha,\beta}(m,n) l_1\left(\mathbf{r}_{int}^* + \begin{bmatrix} 0 \\ 0 \\ mp_s \\ np_t \end{bmatrix}\right). \quad (16)$$

It is noted that the light rays on both sides of Eq. (16) actually reach the image sensor. Therefore, we can apply the technique of light field projection, which is mentioned in Eq. (9) to Eq. (16). The final result specifies the motion compensation model under the fractional ray-space motion:

$$i_2(\mathbf{x}) = \sum_{m=-M}^{M} \sum_{n=-N}^{N} \omega_{\alpha,\beta}(m,n) i_1\left(\mathbf{x} + \begin{bmatrix} (\Delta k_s + m)P_x \\ (\Delta k_t + n)P_y \end{bmatrix}\right). \quad (17)$$

### C. Comparison with conventional motion compensation model

Before proceeding to the next section, it is worth looking back at the conventional motion compensation model that is currently used in most video coding techniques. By referring to prior methods, we can definitely differentiate our model from the existing motion compensation model. The basic concept underlying the conventional motion compensation model is the *pixel brightness constant assumption* [68], which is also used to extract the optical flow [69] from a video sequence. The assumption states that, during a very small transition, the pixel brightness will remain the same:

$$i_2(\mathbf{x}) = i_1(\mathbf{x} + \Delta \mathbf{x}), \quad (18)$$

where $\Delta \mathbf{x} = (\Delta x, \Delta y)$ is the conventional motion vector indicating the pixel displacement $(\Delta x, \Delta y)$ relation between the reference picture $i_1(.)$ and the current picture $i_2(.)$. According to the conventional motion compensation model described in Eq. (18), the motion vector $\Delta \mathbf{x}$ is normally expressed in terms of a function of pixel position $\mathbf{x}$, that is $f: R^2 \to R^2$, such that $\Delta \mathbf{x} = f(\mathbf{x})$. A common realization of the function $f(\mathbf{x})$ is the linear form:

$$f\left(\mathbf{x} = \begin{bmatrix} x \\ y \end{bmatrix}\right) = \begin{cases} a_1 x + b_1 y + c_1 \\ a_2 x + b_2 y + c_2 \end{cases}, \quad (19)$$

where $a_i, b_i, c_i \in R$ (for $i = 1, 2$) are its parameters. In general, it models the affine motion [70]. If $a_i = b_i = 0$, the motion is deduced to translation, which has been widely used in the translational block-based motion compensation [64] of MPEG-1 and MPEG-2, H.264/AVC, H.265/HEVC, and so on. The affine motion was recently adopted in the H.266/VVC by noting its additional benefit of compression performance beyond the translational motion [71]. When the motion vector $\Delta \mathbf{x}$ has fractional precision, that is, sub-pel accuracy, interpolation is executed on the *sensor plane*. This process is referred to as fractional motion compensation [72].

The conventional motion vector is defined as the pixel displacement at the sensor plane. However, our ray motion vector is defined as the ray displacement at the MLA plane by using the concept of the proposed ray-space motion. This key difference leads to significant changes in the motion compensation models, as derived in Eq. (12) and Eq. (17), depending on the integer or fractional ray-space motions. In the integer ray-space motion, since we can observe the pixel displacement, the motion compensation model (Eq. (12)) looks quite similar to the conventional model where the pixel displacement $\Delta \mathbf{x}$ is related to the ray displacement

$(\Delta k_s, \Delta k_t)$ as $\Delta \mathbf{x} = (\Delta k_s P_x, \Delta k_t P_y)$. However, in the fractional ray-space motion compensation, unlike the conventional method (which interpolates the pixel values at the sensor plane to perform the fractional motion compensation), we interpolate the rays that are blocked at the MLA plane and then subsequently derive a new motion compensation model, as shown in Eq. (17). In the next section, we explain how to use the proposed motion compensation model in general video encoders, forming a new motion-compensated prediction.

## V. Light Field Motion-Compensated Prediction

### A. Light field motion-compensated prediction

So far, we have transformed scene point motion into ray-space motion using a ray displacement vector $(\Delta s, \Delta t)$. Subsequently, we derived the corresponding motion compensation model for two specific cases of ray-space motion: integer and fractional ray-space motions. According to Fig. 3, a moving scene point lets a group of pixels also move together at the sensor plane. In this case, once we know the ray displacement $(\Delta s, \Delta t)$, it is possible to predict the pixel values of that group using the proposed motion compensation model. In Fig. 6, we present a diagram of a general video coding framework with the proposed method (highlighted) for motion compensation.

For a given ray displacement vector $(\Delta s, \Delta t)$, we can decompose it into its integer part $(\Delta k_s, \Delta k_t)$ and fractional part $(\alpha, \beta)$, as follows:

$$\begin{cases} \Delta s = (\Delta k_s + \alpha) p_s \\ \Delta t = (\Delta k_t + \beta) p_t \end{cases}. \quad (20)$$

Here, $0 \leq \alpha, \beta < 1$. Clearly, the integer ray-space motion occurs when $\alpha = \beta = 0$, while the fractional ray-space motion occurs otherwise. From this clarification, depending on $\alpha$ and $\beta$, we can decide on a suitable prediction model of either Eq. (12) or Eq. (17) for proper integer or fractional ray-space motions, respectively. We refer to this prediction as *the light field motion-compensated prediction* for a given ray displacement vector. Now, the main task is to estimate the ray displacement vector.

To quickly find the integer part $(\Delta k_s, \Delta k_t)$, we temporarily set $\alpha = \beta = 0$, and the motion compensation model is Eq. (12). An optimization problem should be solved to determine the best $(\Delta k_s, \Delta k_t)$ that minimizes the cost function $E_{\text{int}}(\Delta k_s, \Delta k_t)$:

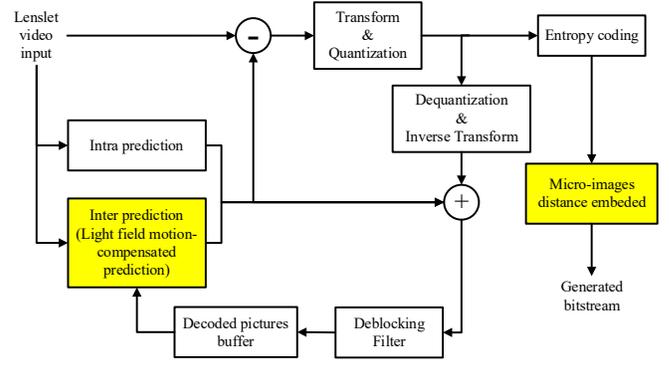

Fig. 6. The diagram of the general video coding framework (highlighted blocks are newly proposed components).

$$\underset{\Delta k_s, \Delta k_t \in Z}{\operatorname{argmin}} E_{\text{int}}(\Delta k_s, \Delta k_t)$$

$$\text{where } E_{\text{int}}(\Delta k_s, \Delta k_t) = \sum_{\mathbf{x} \in PU} \left| i_2(\mathbf{x}) - i_1\left(\mathbf{x} + \begin{bmatrix} \Delta k_s P_x \\ \Delta k_t P_y \end{bmatrix}\right) \right| \quad (21)$$

for all pixels $\mathbf{x}$ in the current coding block (CU) or prediction unit (PU). $|.|$ denotes the absolute operator. From the implementational aspect, the optimization process is equivalent to the conventional motion search where the searching points are located sparsely at multiples of distances $P_x, P_y$ apart along the horizontal and vertical directions at the sensor plane. Therefore, fast search patterns, such as MCP Search [35], are understood quite straightforwardly. We note that although the theoretical analysis behind MCP Search is not given in [35], which just provided statistical analysis by comparing the similarity of the proposed search pattern to that of the full search, our analysis here explains very persuasively why the MCP search [35] performs very well for the lenslet plenoptic video. Additionally, while the motion vector in [35] is still understood as a pixel displacement $(\Delta k_s P_x, \Delta k_t P_y)$ in the conventional way, our integer ray motion vector is understood more accurately by enlightening its relationship with ray displacement $(\Delta k_s, \Delta k_t)$. In the view of data compression, the integer ray motion vector $(\Delta k_s, \Delta k_t)$ can be seen as a quantized version of the conventional motion vector $(\Delta k_s P_x, \Delta k_t P_y)$ with the quantization step $(P_x, P_y)$. The quantization helps to reduce the bit consumption of motion coding data. It is seen that, for integer ray-space motion, while we deliver the same prediction quality as the conventional motion-compensated prediction, the lower motion coding overhead can reduce the bit rate. This leads to better performance in terms of the rate distortion. We will present the compression performance of integer-ray motion in the next section (in Table V) to show the superior performance compared to the conventional method.

After obtaining the integer part $(\Delta k_s, \Delta k_t)$, the remaining issue is to determine the fractional part $(\alpha, \beta)$. Here, the motion compensation model follows Eq. (17). Therefore, we set up another optimization problem that determines the best $(\alpha,\beta)$ that minimizes the cost function $E_{\text{frac}}(\alpha,\beta)$:

$$\operatorname*{argmin}_{0\leq \alpha,\beta<1} E_{\text{frac}}(\alpha,\beta)$$
$$\text{where } E_{\text{frac}}(\alpha,\beta) = \sum_{\mathbf{x}\in PU} |i_2(\mathbf{x}) - \hat{i}_2(\mathbf{x},\alpha,\beta)|. \quad (22)$$

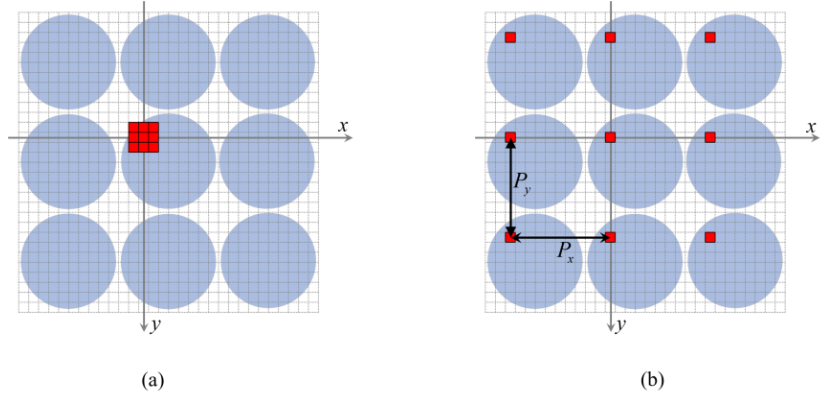

Fig. 7. Demonstration of input samples selection (marked as red dots) for: (a) FIR filter used in conventional fractional motion compensation and (b) our filter used in fractional ray-space motion compensation, as given by Eq. (23). In this figure, for illustrative purposes only, the filter has a size of $3\times 3$, but it could be much larger in actual applications.

Here, $\mathbf{x}$ is the integer pixel position determined by solving Eq. (21), and $\hat{i}_2(\mathbf{x},\alpha,\beta)$ equals:

$$\hat{i}_2(\mathbf{x},\alpha,\beta) = \sum_{m=-M}^{M}\sum_{n=-N}^{N} \omega_{\alpha,\beta}(m,n) i_1\left(\mathbf{x}+\begin{bmatrix}(\Delta k_s+m)P_x\\(\Delta k_t+n)P_y\end{bmatrix}\right), \quad (23)$$

which interpolates the value of pixel $\mathbf{x}$ based on the fractional part $(\alpha,\beta)$. Considering both the prediction accuracy and computational complexity, we decide to check $(\alpha,\beta)$ at the quarter-precision point at the *st*-plane, that is, the position with multiples of a quarter microlens distance:

$$\alpha,\beta \in \left(0, \frac{1}{4}, \frac{2}{4}, \frac{3}{4}\right). \quad (24)$$

Regarding the filter coefficient $\omega_{\alpha,\beta}$, we use the separable filter design method [62] where the 2D filter coefficients are constructed from the product of two individual 1D interpolation filters:

$$\omega_{\alpha,\beta} = \omega_\alpha \cdot \omega_\beta^T, \quad (25)$$

where $\omega_\alpha, \omega_\beta$ are 1D filter coefficients in a vector form. Here, $\omega_\alpha, \omega_\beta$ can be taken from the pre-defined HEVC filter coefficients [63], as shown in Table II. The interpolation filtering in Eq. (23) computes a weighted summation of input samples separated by a step size $(P_x,P_y)$, as seen in Fig. 7(b), explicitly showing the key difference from the conventional interpolation (as seen in Fig. 7(a)). The conventional motion model always takes a few of the immediately neighboring samples for the sub-pel accurate motion compensation [64].

Table II. The 1D filter coefficients used in this paper

| $\alpha$ | $\omega_\alpha(m)$ |
|---|---|
| 0 | {0, 0, 0, 0, 64, 0, 0, 0, 0}/64 |
| 1/4 | {0, -1, 4, -10, 58, 17, -5, 1, 0}/64 |
| 2/4 | {-1, 4, -11, 40, 40, -11, 4, -1}/64 |
| 3/4 | {0, 1, -5, 17, 58, -10, 4, -1, 0}/64 |

The displacement vector describing the ray-space motion with the proposed model can be transmitted to the decoder straightforwardly by making it integer-precise and sending it in the same way as the traditional motion vector. Since we assume quarter-pel accuracy of the displacement vector in this paper, the ray motion vector is scaled before transmission as:

$$\begin{cases} d_s = 4\times(\Delta k_s + \alpha) \\ d_t = 4\times(\Delta k_t + \beta) \end{cases}. \quad (26)$$

The decoder extracts the integer part and fractional part of the ray motion vector as follows:

$$\begin{cases} \Delta k_s = floor(d_s/4) \\ \Delta k_t = floor(d_t/4) \end{cases}; \quad \begin{cases} \alpha = (d_s - 4\Delta k_s)/4 \\ \beta = (d_t - 4\Delta k_t)/4 \end{cases}. \quad (27)$$

In the proposed motion compensation, the decoder should also know the micro-image distance $(P_x, P_y)$. Thus, it is sent over the bitstream header. The Exp-Golomb code [64] is used to binarize the information of both the micro-image distance and ray motion vector.

B. Cooperation with other coding tools

The ray displacement vector is assigned for each PU, which is a basic unit for motion compensation. This information can be treated similarly to a traditional motion vector. Thus, we can re-use existing motion vector coding tools (e.g., advanced motion vector prediction, merge mode, skip mode [64]) to

Table III. Summary of coding tool/methods used in the comparison

| Method | Input format | Configuration |
|---|---|---|
| HEVC | Lenslet | The HEVC screen content profile [65] with all default configurations except the IBC |
| LF-MVC | Multiview | The MV-HEVC (HEVC multiview extension) [34] using the LF-MVC prediction structure in Fig. 8 |
| IBC | Lenslet | HEVC + IBC coding tool [14] |
| MCP Search | Lenslet | HEVC + MCP Search method [35] |
| Proposed | Lenslet | HEVC + proposed ray-space-motion-compensated prediction |
| IBC+ Proposed | Lenslet | HEVC + IBC + Proposed |

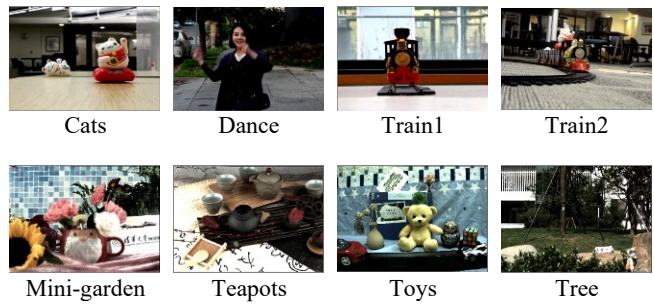

Cats   Dance   Train1   Train2

Mini-garden   Teapots   Toys   Tree

Fig. 9. The dataset for the experiment.

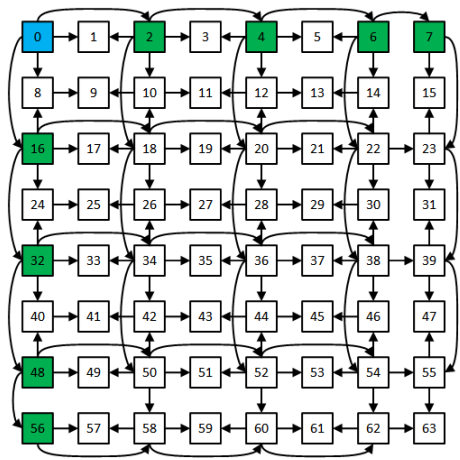

Fig. 8. The LF-MVC prediction structure [28].

transmit the ray motion vector efficiently. Just in case the proposed method is added to some codecs adopting the intra block copy (IBC) tool [14], there can be two kinds of motion vectors: our ray motion vector and the IBC block vector. In this case, the decoder can differentiate whether a received motion vector is for the IBC mode or not by simply checking the reference index. If the reference index refers to the current picture, then the transmitted motion vector should be understood as the block vector of the IBC mode, and if not, it is our ray motion vector. Note that we do not apply the proposed ray motion to the IBC tool since the proposed method only works over the temporal domain. By making these considerations, we can ensure that the proposed method does not conflict with any other existing coding tools.

## VI. EXPERIMENTAL RESULTS AND DISCUSSION

### A. Test condition

To straightforwardly demonstrate the performance of the proposed light field motion-compensated prediction, we implement the method on top of the HEVC reference software (HM-16.21+SCM-8.8) [73] under the screen-content profile [65] with the "*Low Delay B*" configuration [74]. Here, the screen content profile is chosen as a benchmark of the lenslet coding since it is equipped with useful coding tools (e.g., *IBC*, *MotionVectorResolutionControl*) for light field coding [44], while the "*Low Delay B*" configuration is chosen to follow the realistic scenario of low delay light field video streaming and low delay camera storage applications. To generate the 4D light field data for the experiment, the datasets in Fig. 9 are gathered from [44] [75] and are passed through some pre-processing steps [66]. For example, in the case of raw datasets as in [44], the vignetting effect is removed and the hexagonal MLA structure is re-mapped to the rectangular MLA structure. Since the number of viewpoints (i.e., angular resolution) are not the same in the datasets and some viewpoints are missing at the corners (datasets [44]), we decide to crop the angular resolution to 8x8 and convert it to a synthetic lenslet image that has a micro-image distance of $P_x = P_y = 8$. The synthetic lenslet images are then converted to the YCbCr 420 video format so that they can be used as inputs to the HEVC encoder. We take the first 60 frames of each sequence and encode them four times with quantization parameters QPs = {24, 30, 36, 42}. Following the MPEG common test condition [44], the average PSNR (in dB) of the luma channels of all decoded views are computed to measure the plenoptic video quality.

The Bjontegaard delta metric [76] (BD-Rate) is used to evaluate the compression performance of the proposed method and several existing anchors in Table III, including HEVC; LF-MVC, which compresses plenoptic video using the HEVC multiview extension (MV-HEVC) with the *LF-MVC* [28] inter-view prediction structure (see Fig. 8); the macropixel-constrained collocated position (*MCP*) Search [35], which proposes a sensor-based motion searching pattern for the plenoptic video encoder; and the *IBC* mode [14], which exploits the high similarity of the current block and neighboring blocks in the same lenslet image.

### B. Performance evaluation

#### 1) Compression performance

We compare the compression performance of the *proposed method* to those of relevant methods in Table IV. Overall, the proposed method outperforms *HEVC, LF-MVC, IBC,* and *MCP Search* by an average of 18.05%, 49.09%, 16.45%, and 17.44%, respectively. This shows that knowledge of the ray-space motion model can improve the compression performance substantially. Besides, the massive performance gap between the *proposed method* (using the lenslet format as the input) and *LF-MVC* (using the multiview format as the input) reveals the advantage of treating plenoptic data in the lenslet format. For instance, our method is far better than *LF-MVC* for the sequence *Toys*, showing a 93.6% improvement in the BD-Rate performance. Moreover, the *proposed method* does not undermine the coding benefits of the existing plenoptic coding tools, such as *IBC*. To verify this, the *IBC*, one of well-known plenoptic image coding tools, is chosen in this paper to demonstrate the combination. As can be seen from Table IV, combining our method with *IBC*, namely, *IBC + proposed*, achieves 19.63%, 50.11%, 18.08%, and 19.04% gains in the BD-Rate as compared to *HEVC, LF-MVC, IBC,* and *MCP Search,* respectively. We see that the performance increase from the case without *IBC* is not significant, but improves steadily. The rate-distortion curves shown in Fig. 10 reveal substantial performance gains of the proposed method over other existing methods. *LF-MVC* turns out to be the least effective compression scheme, while there are almost no considerable differences among *HEVC, IBC,* and *MCP Search*. However, the *proposed* and *IBC + proposed* methods completely outperform the others, from low-rate to high-rate conditions. The increased performance advantages at higher rates suggest that high quality/fidelity lenslet videos are more manageable.

Table IV. BD-Rate performance of the proposed method with various anchors

| Test Sequences | Proposed vs. HEVC | Proposed vs. LF-MVC | Proposed vs. IBC | Proposed vs. MCP Search | IBC + Proposed vs. HEVC | IBC + Proposed vs. LF-MVC | IBC + Proposed vs. IBC | IBC + Proposed vs. MCP Search |
|---|---|---|---|---|---|---|---|---|
| Cats | -15.0% | -22.5% | -11.4% | -13.1% | -18.2% | -25.6% | -14.7% | -16.3% |
| Dance | -23.6% | -45.2% | -23.6% | -23.4% | -23.6% | -45.2% | -23.5% | -23.3% |
| Train1 | **-27.3%** | -20.3% | **-25.6%** | **-26.6%** | **-29.1%** | -22.4% | **-27.5%** | **-28.4%** |
| Train2 | -24.2% | -31.6% | -23.6% | -23.8% | -24.9% | -32.2% | -24.4% | -24.8% |
| Mini-garden | -16.5% | -36.9% | -15.1% | -15.7% | -17.9% | -38.1% | -16.5% | -17.1% |
| Teapots | -19.0% | -60.9% | -17.7% | -18.2% | -20.2% | -61.6% | -18.9% | -19.4% |
| Toys | -8.0% | **-93.6%** | -4.9% | -7.8% | -11.1% | **-93.8%** | -8.2% | -10.9% |
| Tree | -10.8% | -81.7% | -9.7% | -10.9% | -12.0% | -82.0% | -10.9% | -12.1% |
| **Average** | **-18.05%** | **-49.09%** | **-16.45%** | **-17.44%** | **-19.63%** | **-50.11%** | **-18.08%** | **-19.04%** |

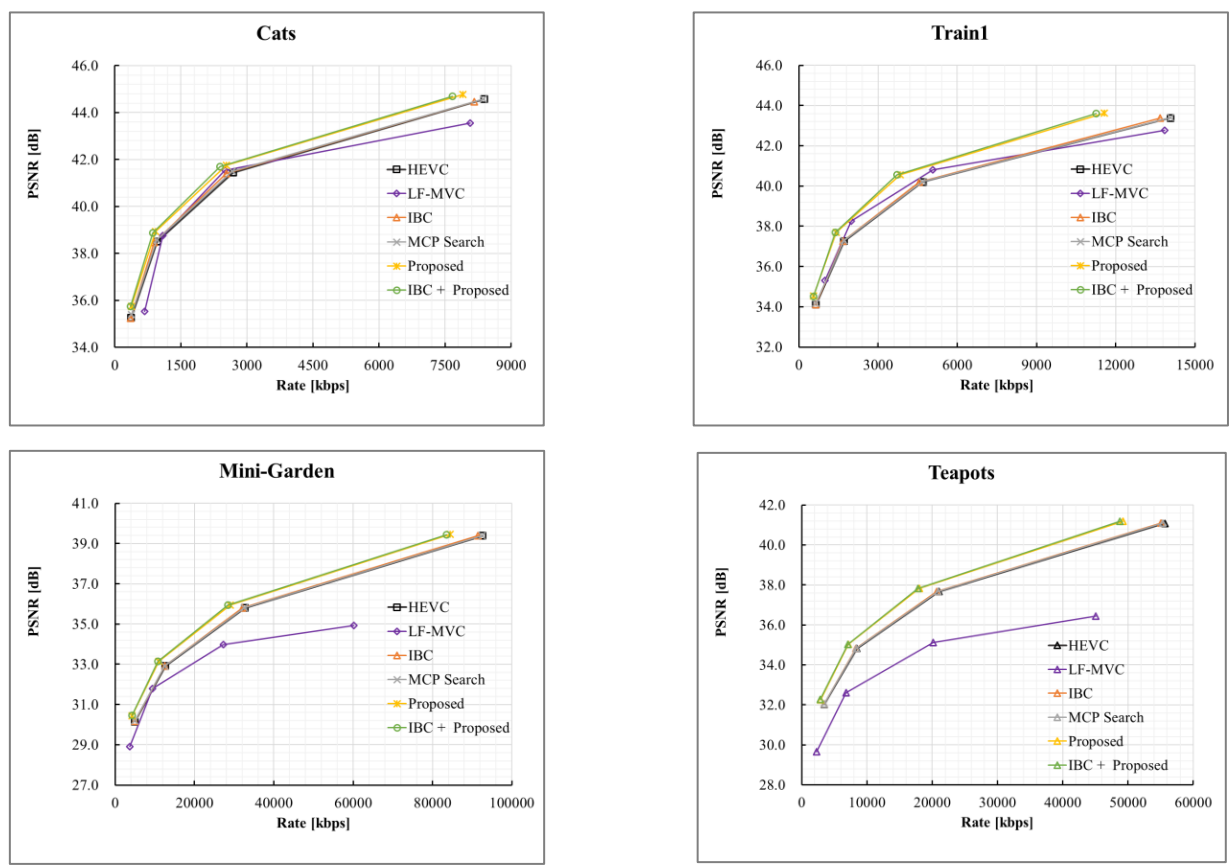

Fig. 10. Comparison of rate-distortion performance.

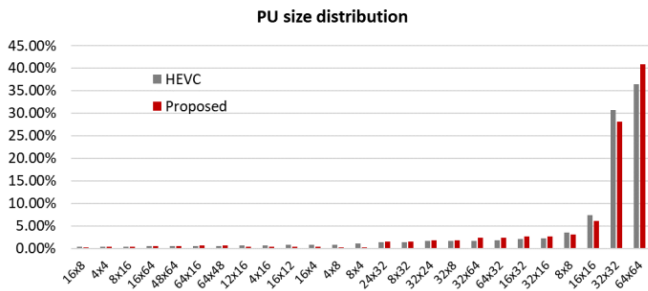

Fig. 11. Comparison of PU size distribution (*Train1* at QP=30).

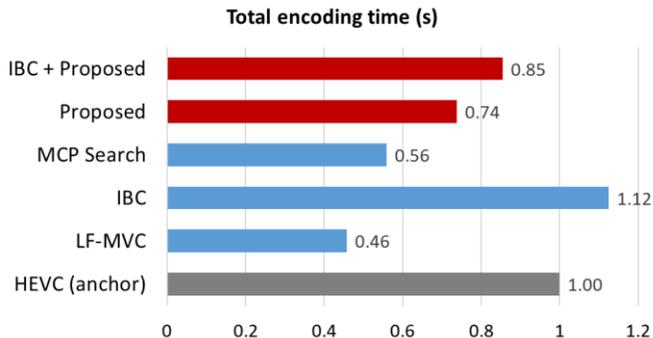

Fig. 12. Comparison of relative encoding time-complexity with HEVC as an anchor.

Table V. The effect of additional ray motion precision on the compression performance (anchor is the HEVC; note that our proposed method estimates ray displacement at quarter-precision).

| Ray motion<br>Test sequences | Integer-precision | Half-precision | Quarter-precision |
|---|---|---|---|
| Cats | -7.0% | -12.3% | -15.0% |
| Dance | -5.8% | -19.2% | -23.6% |
| Train1 | -4.8% | -22.8% | -27.3% |
| Train2 | -5.3% | -20.8% | -24.2% |
| Mini-garden | -7.9% | -13.3% | -16.5% |
| Teapots | -10.9% | -14.2% | -19.0% |
| Toys | -3.9% | -6.0% | -8.0% |
| Tree | +4.9% | -2.5% | -10.8% |
| **Average** | **-5.09%** | **-13.89%** | **-18.05%** |

The PU size distribution in Fig. 11 also supports the effectiveness of the proposed method. Compared to *HEVC*, more PUs end up with larger sizes when using the proposed method. This indicates better prediction at lower signaling costs.

We further study the importance of estimation precision of the ray displacement by investigating following two extra cases: *integer-precision,* fixed fractional part, $\alpha = \beta = 0$ ; *half-precision,* $\alpha, \beta \in \{0; 2/4\}$ in additional to the *quarter-precision*, $\alpha, \beta \in \{0; 1/4; 2/4; 3/4\}$. The results are given in Table V, in which a higher precision of the ray motion is seen to lead to better coding performance. The compression performance rapidly grows from integer-precision to half-precision, but slowly increases when further moving to quarter-precision. Although a higher precision should give more accurate prediction, its consumption of more bits to transmit the ray displacement information leads to a trade-off in the rate-distortion performance. That could be the main reason for the slow performance increase at the quarter-precision.

*2) Encoding time-complexity*

For each method we have tested, we also evaluate the encoding time-complexity by measuring the total encoding time (s). These experiments are conducted using a single PC i7-8700 CPU @ 3.2GHz, 32GB RAM with 64-bit Windows 10 installed. In Fig. 12, we compare the relative time-complexity with the anchor (*HEVC*). *LF-MVC* is seen to be the most light-weight compression method. However, it also has the lowest compression efficiency. Turning on the *IBC* tool increases the time-complexity to 112% due to the exhaustive search of similar blocks in the same picture. Apart from that, *MCP Search* reduces the encoding time to 56%, mainly because its searching pattern is optimized for plenoptic video, enabling suitable motion vectors to be found with less searching. Remarkably, the *proposed* and the *IBC + proposed* are even faster than the anchor, featuring time-complexity values that were 74% and 85% those of HEVC, respectively. The reason for this is that a similar process as *MCP Search* is executed by the proposed methods to find the integer part of the ray motion vector (refer to Eq. (21)); thus, the *proposed methods* also inherit the low complexity of *MCP Search*. A noticeable time difference between the *proposed* and *MCP Search* is that our implementation of interpolation filtering shown in Eq. (23) is not yet well-optimized as HEVC reference software [73] which adopts the SIMD vectorization [78]. It is seen that the proposed scheme not only boosts the compression performance but also significantly reduces the time-complexity, both of which give benefits to the practical design and applications.

## VII. Conclusion

Our understanding of the ray-space motion model proposed in this paper led us to design a new motion-compensation method suitable for plenoptic video in the lenslet format. It is shown to provide drastic compression improvement at a lower encoding time. However, the proposed model has a simplifying assumption where it does not consider *z*-motion. Therefore, it might not perfectly express all motions in reality. In our future work, we would like to address this issue to further improve the model.

## APPENDIX A

Proof of constant ray displacement for all *uv* positions under no Z-motion of the scene point/image point.

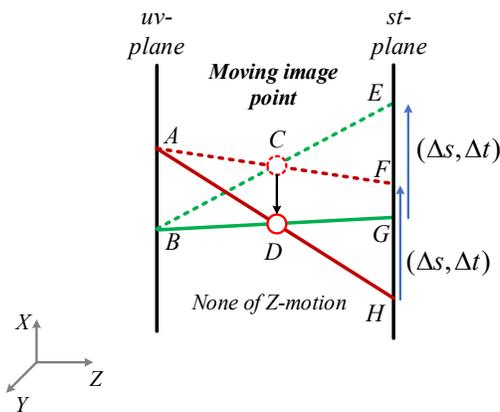

Fig. 13. Simplified ray-space motion model with intersections marked by capital letters. Here, to focus on the problem, the scene point motion is not drawn.

In Fig. 13, we would like to prove that the displacement *GE* is equal to *HF*. First, under the no Z-motion assumption, the movement *C* to *D* should be parallel to the *st*-plane; thus, triangles $\Delta BEG \sim \Delta BCD$ and $\Delta AFH \sim \Delta ACD$. From these similar triangles, we have:

$$\begin{cases} \dfrac{BG}{BD} = \dfrac{GE}{DC} \\ \dfrac{AH}{AD} = \dfrac{HF}{DC} \end{cases} \Rightarrow \begin{cases} \dfrac{BG}{BD} - 1 = \dfrac{GE}{DC} - 1 \\ \dfrac{AH}{AD} - 1 = \dfrac{HF}{DC} - 1 \end{cases}$$
$$\Rightarrow \begin{cases} \dfrac{DG}{BD} = \dfrac{GE - DC}{DC} \\ \dfrac{DH}{AD} = \dfrac{HF - DC}{DC} \end{cases} \qquad (28)$$

Also, triangles $\Delta DHG \sim \Delta DAB$, implying that:

$$\dfrac{DG}{BD} = \dfrac{DH}{AD}. \qquad (29)$$

Combining Eq. (28) and Eq. (29), it turns out that:

$$\dfrac{GE - DC}{DC} = \dfrac{HF - DC}{DC} \Rightarrow GE = HF. \qquad (30)$$

Therefore, the proof is concluded. ∎